\documentclass[conference]{IEEEtran}

\usepackage{cite}

\ifCLASSINFOpdf
\else
\fi
\usepackage{url}

\usepackage[english]{babel}

\makeatletter
  \@addtoreset{paragraph}{section}
\makeatother

\usepackage[hyperfootnotes=false]{hyperref}
\hypersetup{
  pdfinfo={
    Title={Reliable Low Latency Wireless Communication Enabling Industrial Mobile Control and Safety Applications},
    Author={Sergiy Melnyk},
  }
}

\usepackage[dvips]{graphicx}
\graphicspath{{figures/}}

\usepackage{xcolor}

\usepackage{soul}

\usepackage[textsize=scriptsize]{todonotes}

\usepackage{cleveref}

\usepackage{siunitx}
\DeclareSIUnit[qualifier-mode = combine]{\dBm}{\deci\bel\of{m}}
\DeclareSIUnit[qualifier-mode = combine]{\dBi}{\deci\bel\of{i}}

\usepackage{comment}

\usepackage{subfig}
\usepackage{pgfplots}
\usetikzlibrary{plotmarks}
\usepackage{tikz,pgfplots}
\usepackage{multirow}
\usepackage{standalone}
\usepackage{amsfonts}

\begin{document}
%

\title{Reliable Low Latency Wireless Communication Enabling Industrial Mobile Control and Safety Applications}



%
\author{
  \IEEEauthorblockN{
    Sergiy Melnyk\IEEEauthorrefmark{1},
    Abraham Gebru Tesfay\IEEEauthorrefmark{1},
    Khurshid Alam\IEEEauthorrefmark{1},
    Hans D. Schotten\IEEEauthorrefmark{1},\\
    Vladica Sark\IEEEauthorrefmark{2},
    Nebojsa Maletic\IEEEauthorrefmark{2},
    Mohammed Ramadan\IEEEauthorrefmark{2},
    Marcus Ehrig\IEEEauthorrefmark{2},\\
    Thomas Augustin\IEEEauthorrefmark{3},
    Norman Franchi\IEEEauthorrefmark{3} and
    Gerhard Fettweis\IEEEauthorrefmark{3},
  }
  \IEEEauthorblockA{
    \IEEEauthorrefmark{1}
    Intelligent Networks,
    German Research Center for Artificial Intelligence (DFKI GmbH) \\
    Email: \{sergiy.melnyk,\,abraham\_gebru.tesfay,\,khurshid.alam,\,hans\_dieter.schotten\}%
    @dfki.de
  }
  \IEEEauthorblockA{    
    \IEEEauthorrefmark{2}
    IHP, Frankfurt (Oder), Germany\\
    Email: \{sark,\,maletic,\,ramadan,\,ehrig\}@ihp-microelectronics.com
  }
  \IEEEauthorblockA{    
    \IEEEauthorrefmark{3}
    Vodafone Chair Mobile Communications Systems,
    Technische Universit\"at Dresden, Germany\\
    Email: \{thomas.augustin,\,norman.franchi,\,gerhard.fettweis\}@tu-dresden.de
  }
}

\maketitle

\begin{abstract}
Advanced industrial applications for human-machine interaction such as augmented reality support for maintenance works or mobile control panels for operating production facility set high demands on underlying wireless connectivity solution. Based on 802.11 standard, this paper proposes a concept of a new system, which is capable of those requirements. For increasing reliability, an agile triple-band (\SIlist{2.4; 5; 60}{\GHz}) communication system can be used. In order to deal with latency and deterministic channel access, PHY and MAC techniques such as new waveforms or hybrid MAC schemes are investigated. Integration of precise localization introduces new possibilities for safety-critical applications.  
\end{abstract}


\section{Introduction}

\let\thefootnote\relax\footnote{\textit{\textbf{Submitted to 3th VDE/ITG Conference on Mobile Communication (23. VDE/ITG Fachtagung Mobilkommunikation), Osnabr\"uck, Mai 2018}}}%
Shifting manufacturing facilities to cyber-physical production systems (CPPS) provides new opportunities to the industry for organising and controlling the manufacturing process. However, only by using wireless communication systems, CPPS becomes flexible and easily scalable. There are several initiatives, which are supporting the developement of Industrial Internet. One of them is the German initiative ``Industrie 4.0'' \cite{aktas:2017}.

The majority of current activities are considering process or factory automation. Closed-loop applications set advanced requirements on wireless solutions. Besides the flexibility and reliability,  communication latency below \SI{1}{\ms} is required, whereas the data rate is moderate. This makes the development of dedicated systems necessary \cite{bockelmann:2017}.

Another large field of industrial applications, which future development also relies on the quality of wireless connection, is human-machine interaction. By using such techniquess as interactive augmented reality (AR), mobile control or others which are supporting field personnel, productivity of the manufacturing processes could be significantly increased.  
On the one hand, HMI applications have moderate requiremets on latency as compared to automation applications. On the other hand, they demand high data rates in order to e.\,g realise offloading of comprehensive video processing or high link reliability for implementing safety protocols over wireless \cite{melnyk:2017}.

Most of established industrial wireless solutions, such as WirelessHART, primarily address automation applications. Besides, solutions known from home and office sector as for example Bluetooth or WLAN are about to find their way into the manufacturing area  \cite{bregulla:2011}. The latter is also well suited for HMI applications as it is designed to deal with high data rate and heterogeneous traffic. However, it still lacks deterministic channel access or appropriate reliability prohibiting its application in an industrial environment.

In this paper, we investigate some improvements, which could make WLAN capable of industrial HMI support. First, we propose a concept of agile  triple-band communication (\SIlist{2.4; 5; 60}{\GHz}) for increasing reliability by multi-connectivity. As to deal with latency and deterministic channel access, some techniques for physical (PHY) and medium access (MAC) layer such as new waveforms or hybrid MAC schemes are investigated. Last but not least, the integration of precise localization system can introduce new possibilities such as safety zones for safety-critical applications.

The rest of the paper is organised as follows.
Section~\ref{sec:requirements} describes the industrial HMI applications we concentrated on in our work. Further, it presents their requirements on the communication system.
In Section~\ref{sec:system}, we propose a system concept, which would help to fulfil the aforementioned requirements.
Section~\ref{sec:technologies} gives a deep insight into the proposed solutions  and technologies.
Finally, our work is summarised by  Section~\ref{sec:conclusion}.

\section{Applications and Requirements }
\label{sec:requirements}

\begin{figure}[b]
  \centering
  \includegraphics[width=.77\linewidth]{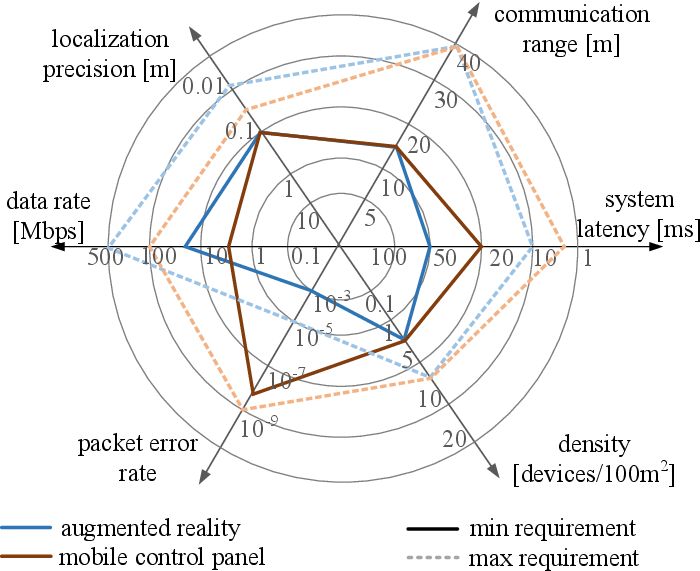}
  \caption{Industrial HMI requirements \cite{melnyk:2017}}
  \label{fig:requirements}
\end{figure}

In our work, we target those future-oriented applications, user-experience and acceptance of which highly depend on the performance of the applied wireless communication solution.
More details are given in the following subsections, whereas  \Cref{fig:requirements} depicts overall requirements.




\subsection{Augmented Reality Applications}

Augmented Reality is a technique of putting additional information (augmentations) to live pictures in realtime~\cite{barfield:2015}. This technique may be applied on a videostream shown on a screen or tablet as well as on a head-up display. Possible applications may be e.\,g. ovelaying service data at a glance or providing service staff with live manuals augmenting next service steps~\cite{gavish:2015:778}.

The core part of any augmented reality application is a stable localization based on video data processing. This method only allows a precise definition of the camera position and orientation with respect to every video frame. Afterwards, augmentations can be integrated to the real world by rendering them on the proper positions. However, greedy processing algorithms cannot be run on mobile devices due to their low computation performance. One possible solution is to offload the computations to a powerfull server \cite{hasper:2014:156}.

In order to avoid the occurance of cyber sickness, end-to-end latency of the offloaded processing steps may not exceed \SI{70}{ms} and in the best case it should stay below \SI{20}{\ms}. Video streaming produces high datarate traffic and depending on the quality of the video stream as well as compression method, the required bandwidth may reach from \SI{500}{\mega\bit \per \s} up to \SI{6.6}{\giga\bit \per \s} \cite{melnyk:2017}.

\subsection{Safety-Critical Applications}

Operation of machines or production units by means of a mobile control panel (MCP) provides flexibility and increases user experience of workers, leading to increased productivity. In order to guarantee the safety of personnel,  mobile devices need to be equipped with safety guard functionalities. In order to fulfil the IEC\,61508 standard on automation safety, a dedicated protocol such as  PROFIsafe or OpenSafety needs to be used on networked communication technologies \cite{sato:2015:299}.

The principle behind those kind of protocols is sending ``Sign-of-Live'' messages from the controller device to programmable logic controller (PLC) of the targeted machine. Hereby, cycle time of \SIrange{4}{8}{\ms} is required. If such kind of message does not arrive during a defined time window (watchdog interval), the machine would perform an emergency stop. Typical watchdog intervals are below \SI{30}{\ms} and for some applications even below \SI{10}{\ms}. This makes safety critical applications also critical to communication delay, i.\,e. latency. In addition, safety protocols are particularly depending on the reliability of the network since losing any safety packet would lead to an emergency stop of the machine. \cite{melnyk:2017}.

\section{Communication System Concept}
\label{sec:system}

Based on our previous investigations of state of the art communication systems \cite{melnyk:2017},  WLAN standard IEEE 802.11\,n/ac  offers the best coverage of the above requirements as compared to other systems. However, WLAN still lacks sufficient data rate, reliability, deterministic channel access etc. Nevertheless, we would like to propose some technologies, which are able to turn WLAN to promising technology for industrial use.
  
In order to overcome the shortcomings of the \SIlist{2.4; 5}{\GHz} frequency bands, communication in \SI{60}{\GHz} band  is introduced. It is able to  handle high data rates due to the high channel bandwidth available. Furthermore, the available bandwidth facilitates high precision of localization setup. We propose to realise an agile  multiband communication system which is able to communicate in any of the \SIlist{2.4; 5; 60}{\GHz} frequency bands. An appropriate communication band is to be chosen based on the requirements of the application. Moreover, several bands can be used for increasing reliability of the communication link due to redundancy.

Integration of a required positioning system into communication hardware can significantly reduce deployment costs and efforts. In this paper, a two way ranging (TWR) approach is proposed which can be implemented for any of the  desired frequency bands and can be alternately used together with data transmission.

Further improvements  on latency and throughput can be achieved by using new flexible waveforms. We suggest the usage of Generalized Frequency Division Multiplexing (GFDM), also we present its PHY realization. It allows, due to its flexibility, to adjust the waveform and frame structure with regard to application requirements.

Last but not least,  deterministic channel access needs to be provided in order to guarantee deadlines of HMI protocols. Here, we suggest the usage of hybrid MAC algorithms providing a  contention free period (CFP) controlled by an access point (AP). Our investigations showed that the usage of appropriate scheduling algorithms lead to a deterministic behavior of the network which is appropriate for safety critical applications.

Detailed results on aforementioned improvements and technologies are depicted in the next section.

\section{Enabling Technologies}
\label{sec:technologies}

In this section, we discuss single technologies of presented system concept. Additionally, we provide our recent research results on these topics.

\subsection{\SI{60}{\GHz} System}

\begin{figure}[b]
  \centering
  \includegraphics[width=.3\linewidth]{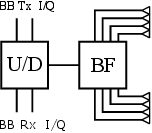}
  \caption{\SI{60}{\GHz} beamforming system}
  \label{fig:60ghz_system}
\end{figure}

The \SI{60}{\GHz} beamforming system developed in Prowilan is shown in Figure~\ref{fig:60ghz_system}. The analog front-end (AFE) consists of a beamforming (BF) integrated circuit (IC), up-conversion and down-conversion ICs.

\begin{figure}[t]
  \centering
  \subfloat[]{%
    \includegraphics[width=.51\linewidth]{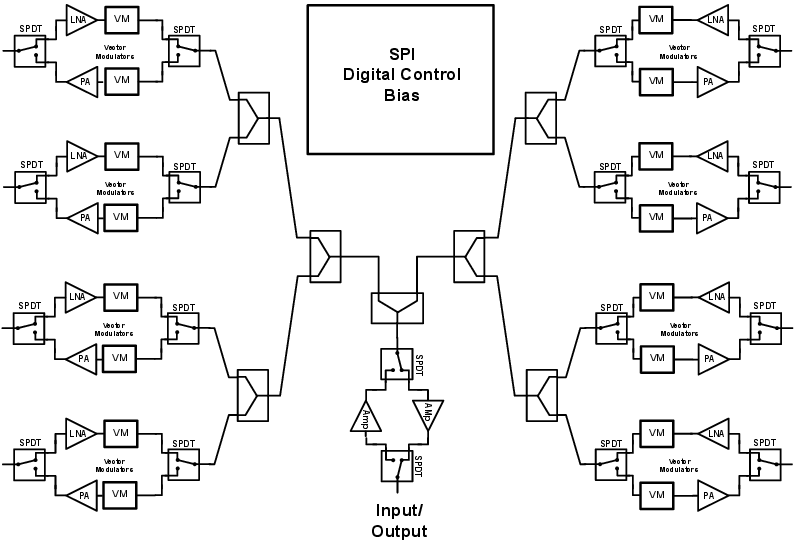}%
    \label{fig:60ghz_beamformer}%
  }%
  \qquad%
  \subfloat[]{%
    \includegraphics[width=.35\linewidth]{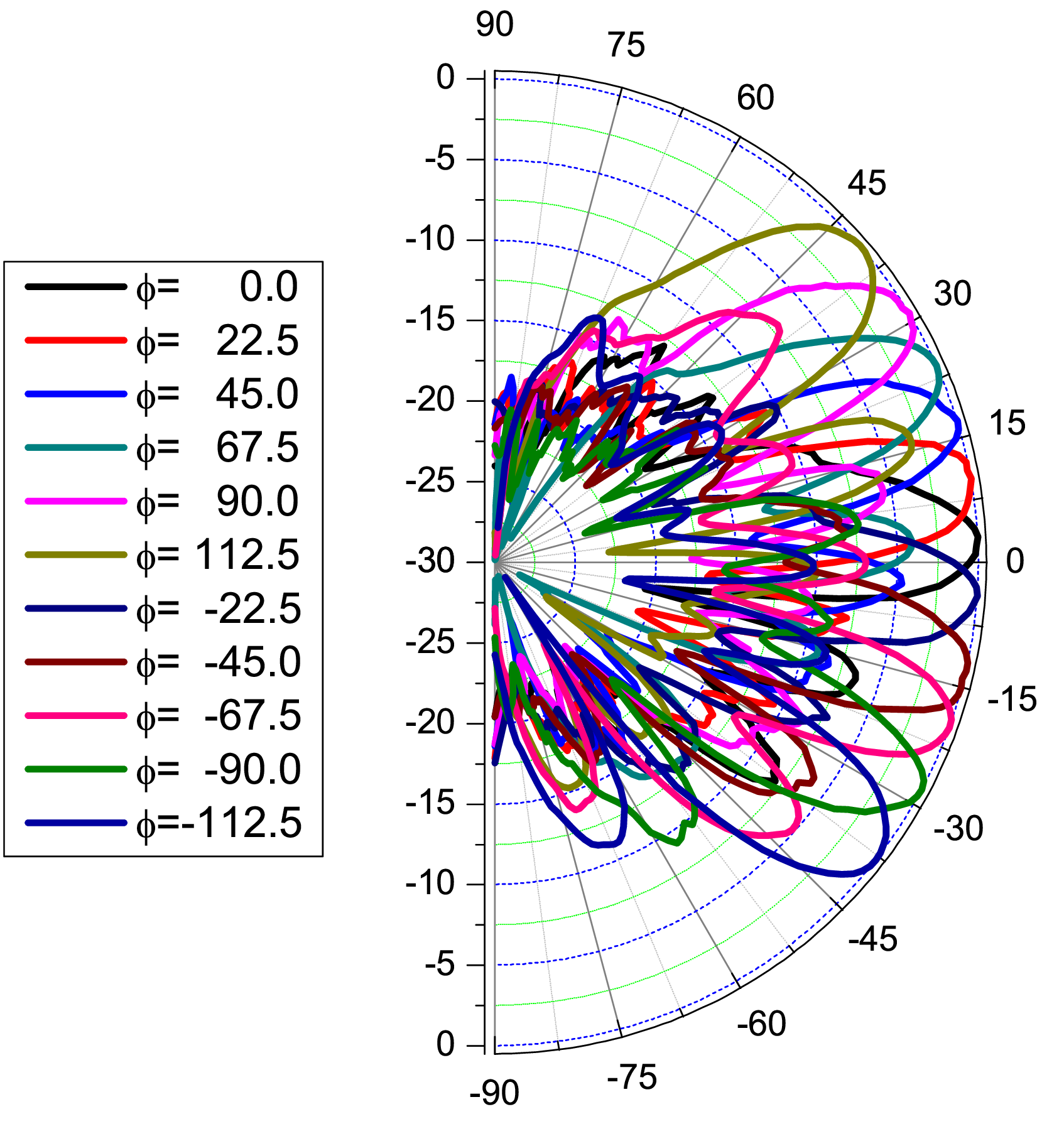}%
    \label{fig:60ghz_meas}%
  }
  \caption{%
    \protect\subref{fig:60ghz_beamformer} Beamformer block diagram; 
    \protect\subref{fig:60ghz_meas} Beamforming measurements%
  }
  \label{fig:60ghz_chip}
\end{figure}

The \SI{60}{\GHz} BF IC is designed in IHP's \SI{130}{\nm} SiGe:C BiCMOS technology \cite{malignaggi:2017:1602} performing beamforming in transmit mode and beam combining in received mode. The chip integrates eight bidirectional channels, each consisting of an antenna switch for Tx/Rx-path selection, a Tx-path with power amplifier and Tx vector modulator and an Rx-path with low-noise amplifier and Rx vector modulator. Additionally, a Tx/Rx-path selection switch and a compensation amplifier are included at the interface to the corresponding \SI{60}{\GHz} up-/down mixer. The Figure~\ref{fig:60ghz_beamformer} shows the chip block diagram. Since RF switches are included in the Tx path and in the Rx path, the IC can be used in TDD systems with one TRx antenna. For the use in a stand-alone Rx or Tx module, the IC can be set to single Tx/Rx operation through a dedicated pin or through the digital control. The chip has an integrated SPI controller, and eight registers for storing eight beams for fast selection in less than \SI{20}{\ns}. More details are given in \cite{malignaggi:2017:1602}. 
The functionality of the beamforming IC is tested on a RF test board having an antenna array with eight dual-patch elements with a max gain around \SI{11}{\dBi}. The results of the beamforming measurement are depicted in Figure~\ref{fig:60ghz_meas}.

\subsection{Flexible GFDM Waveform}

In industrial environments, wireless networks operating in unlicensed bands typically suffer because multiple radio technologies share the same frequency resources, causing cross-technology interference. However, in the industrial scenarios considered in this paper, usually various technologies have to be supported in order to enable connect augmented reality and safety critical applications. One solution to solve those challenges is to use a flexible waveform, which can be adapted to the different requirements of the respective application. Therefore, in this paper, a Generalized Frequency Division
Multiplexing (GFDM) based frame structure \cite{Matthe2016} is investigated, which can be easily adjusted to an application specific waveform, including most state-of-the-art waveforms, in order to fulfill the requirements of industrial services. Another benefit of GFDM, as described in [last paper] is the increase in spectral efficiency due to low out-of-bounds emission. This is because the subcarrier of the GFDM waveform and the communication channels as well can be allocated more tightly together without introducing adjacent channel of sub-channel interference.

\subsubsection{Flexible Phsyical Layer}
As described before, in industrial scenarios, multiple different services with various needs have to be supported at the same time. These services have different requirements regarding data rates, reliability, and latency and usually have to be supported by multiple radio access techniques, including various waveform configurations. In order to avoid interference and channel access competition, advanced techniques like network slicing has to be used, where all RATs are harmonized with each other. However, the synchronization of these different technologies is non-trivial. Additionally, in order to provide network coverage to various industrial applications, different access points for radio access technology  have to be deployed in the area. An approach to reduce this effort is usage of a flexible physical layer (PHY) chip set, which is capable of providing multiple application-specific transmissions. In order to support multiple application at the same time, the physical layer has to be reconfigured quickly to different frame structures and waveforms. Therefore, in \cite{Danneberg} a flexible PHY based on GFDM has been developed whose architecture will be presented in the following section.

\subsubsection{Transceiver Architecture}
The block diagram of flexible GFDM-based PHY in \Cref{phy_pic} shows components needed for the generation of a flexible waveform based on GFDM. First description of the block functionalities has been published already in \cite{7390956} and \cite{Danneberg} and will be shortly summarized. Beside PHY components like channel encoder, QAM-Mapper synchronization, channel estimation which are found in most state-of-the-art implementation, it also consists of special PHY blocks which have been developed to support GFDM-based waveforms. First special block is the resource mapper, which maps complex constellation symbols to a two dimensional resource grid. The resource grid strongly depends on the frame structure and targeted waveform configuration. The modulator can be divided into discrete Fourier transformation and the modulation process itself. The modulator splits the output of inverse discrete Fourier transformation into sub-symbols and samples. Afterwards pulse shaping is performed. Subsequently,  outcomes of all parallel branches are added together. After modulation, additional time window and preamble have to be added. Here, the preamble is used for synchronization and channel estimation. Cyclic prefix and cyclic suffix can be applied on both preamble and data block. The synchronization is based on the Schmidl-Cox algorithm \cite{650240,Gaspar2014}. For estimation, a frequency domain least-square estimator with subsequent linear interpolation is used. The principle of the resource demapper and demodulator is the inverse as for counterpart in the transmitter chain.

\subsubsection{Measurement Results}
\begin{figure}[t]
  \centering
  \includegraphics[width=.7\linewidth]{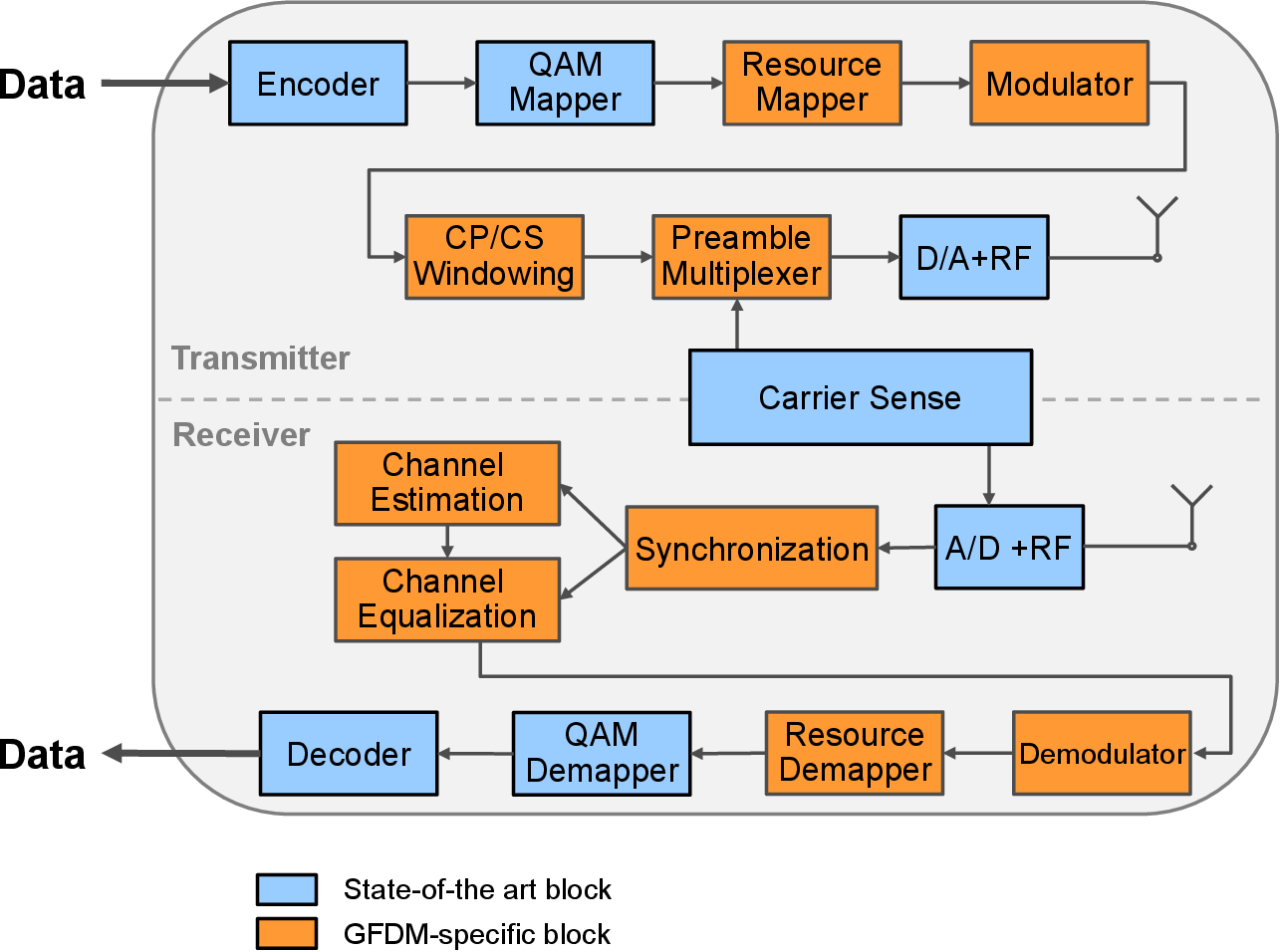}\\
  \caption{Flexible PHY architecture}\label{phy_pic}
\end{figure}

The implemented PHY transceiver architecture of figure \ref{phy_pic} is evaluated in a laboratory environment using a cable setup where latency and performance behavior has been analyzed.
As result of PHY implementation on an FPGA, configuration between different frame structures and waveforms can be switched every 20 clock cycles, which gives enough time to support different services simultaneously. However, depending on the on the waveform configuration, the number of used processing blocks differs and thus required latency varies. Therefore, single-carrier waveform configuration has the lowest latency and in contrast to that GFDM, where all processing blocks of \ref{phy_pic} are used, has the highest latency. Neverthless, for all configurations the latency is always below $100 \mu s$ and therefore fulfills the latency requirements for targeted HMI applications in industry. Evaluation of bit error ratio (BER) performance shows that BER of $10^{-5}$ can be achieved below measured 5dB. This can be further improved by further multi-connectivity concepts on PHY, MAC or higher Layers as presented in MAC-Chapter and \cite{ehrig:2017:1301}.

\subsection{Channel Access}

Channel access strategy is one of the most important components of a wireless system, which among others influences its scalability, reliability and determinism. Channel access method which is mandatory in 802.11 standard  is called distributed coordination function (DCF) and is a realisation of collision avoidance  mechanism. The amendment .11e of the standard further defines enhanced distributed channel access as a part of hybrid coordination function (HCF). It improves DCF by traffic prioritisation. These two techniques are well established in commercial hardware. However, they are known to be non-deterministic due to contention based channel access.

Besides, there exists point coordination function (PCF), which introduses contention free period (CFP). During this period, AP takes the control of the medium and sequentially polls registered stations. Whereas PCF is provided to guarantee time based access, it suffers from unknown transmission time, unpredictable beacon delays and more \cite{mangold:2003:40}. An improvement is provided by HCF controlled channel access (HCCA), which is capable of traffic class differentiation, transmit time control and some more features.
\cite{ehrig:2017:1301}

\subsubsection{Comparisson of Channel Access Methods}

\begin{figure}
  \centering
  \includegraphics[width=.8\linewidth]{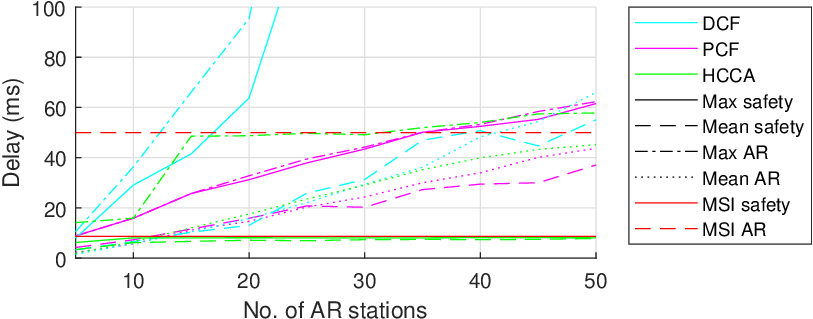}
  \caption{Latency vs. number of AR stations}
  \label{fig:delay}
\end{figure}

In this section, we provide a comparison between DCF, PCF and HCCA with respect to their latency performance. For  out investigation, we created MATLAB models of that three channel access methods. In the case of HCCA, reference scheduler proposed by the standard was used. We differentiated safety and AR stations. Safety stations produce cyclic traffic every \SI{8}{\ms} and have maximum service interval (MSI) of \SI{8}{\ms}. The number of safety stations was fixed to 2. AR stations produce high data rate traffic with MSI of \SI{50}{ms}. The number of stations vary from 5 to 50. For PCF and HCCA, both stations types were scheduled on the CFP. Figure~\ref{fig:delay} shows the simulation results.

As expected, DCF features the worst performance. Whereas the mean delay of safety packets is below MSI for less  AR stations, maximum delay is always above the deadline.  Also, it should be noticed, that maximum latency drastically increases by increasing number of stations due to increasing number of collisions. Therefore, MSI can be reached in average, but it cannot be guaranteed.

For PCF, the performance is slightly better as compared to DCF. However, average delay of safety stations is suitable for less than 12 stations, whereas maximum delay exceeds the MSI for more than 5 stations. In contrast to DCF, this delay can be guaranteed due to polling technique used by PCF. However, both safety and AR traffic show similar delay performance. As PCF does not differentiate traffic classes, safety and AR traffic are treated in the same manner. It means that superframe duration of PCF should be always adopted with respect to lowest MSI, which considerably limits PCF performance for heterogeneous traffic.

In contrast to both other access methods, HCCA can deal with heterogeneous traffic. For safety traffic, average delay almost reaches maximum delay values. Furthrermore, latency never exceeds the MSI. However, the maximum delay of AR traffic exceeds the MSI for more than 30 stations. That is, HCCA cannot serve more than 30 AR stations. Nevertheless, it should be noticed, that safety traffic could be scheduled for any number of AR stations, which significantly overperforms PCF.

\subsubsection{Comparison of HCCA scheduling techniques}

\begin{figure}
  \centering
  \includegraphics[width=.8\linewidth]{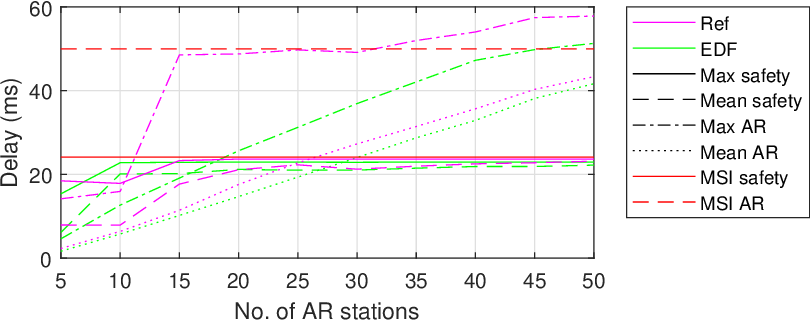}
  \caption{HCCA with reference vs. EDF scheduler}
  \label{fig:scheduler}
\end{figure}

In this section, we provide an investigation of HCCA scheduling techniques. We concider two different schedulers: reference scheduler provided by the standard and Earliest Deadline First (EDF) scheduler \cite{ruscelli:2014:777}. Reference scheduler polls registered stations in round robin fashion, however, it is able to distinguish between traffic classes with different MSIs. A scheduling table is generated once the MSIs are known, and it is updated on significant traffic change only. In contrast, EDF scheduler steadingly update its scheduling table. The packets are sorted based on their deadline. The packet with the closest deadline is also the first to be delivered.

The simulations we carried out are similar to those in previous section. The only  difference is that the MSI of safety traffic was increased to \SI{24}{\ms}. As we can see from Fig.~\ref{fig:scheduler}, both schedulers perform similar with respect to safety traffic. Also, average delay of AR traffic is similar. However, for reference scheduler, maximum delay reaches MSI for already low number of stations due to fixed polling intervals. In contrast, max. delay provided by EDF scheduler rises linearly with number of AR stations. Also, EDF is able to schedule at least 45 AR stations, which significantly overperforms reference scheduler. This result shows the flexibility of EDF scheduler as compared to reference scheduler.

\subsection{Localization }

\begin{figure}[t]
  \centering
  \subfloat[]{\includegraphics[width=0.43\linewidth]{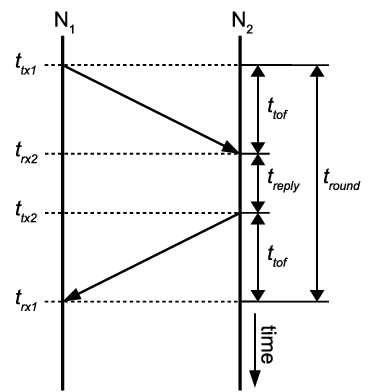}  \label{fig:twr_basic}}
  \quad
  \subfloat[]{\includegraphics[width=0.4\linewidth]{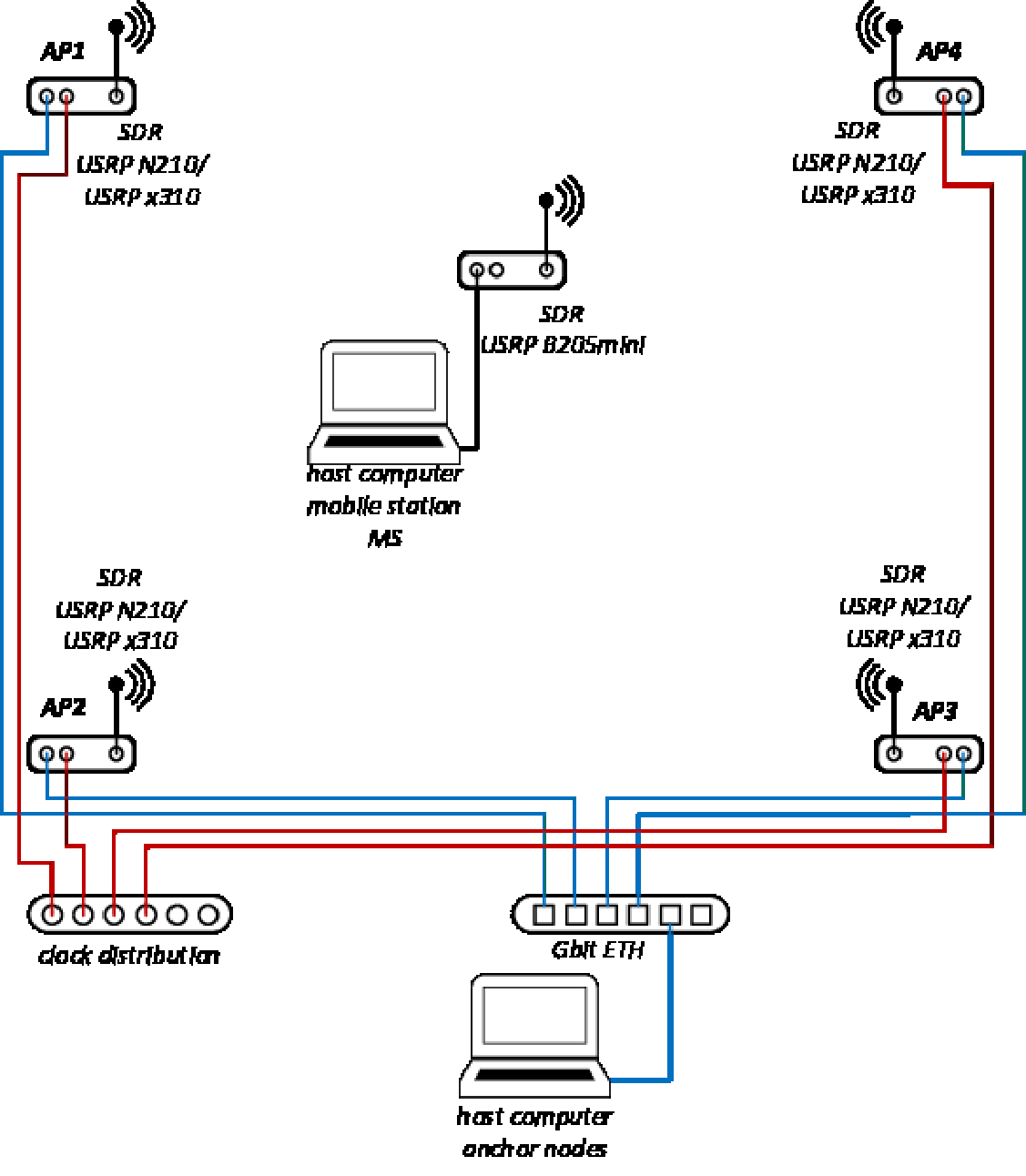}  \label{fig:loc}}
  \caption{
    \protect\subref{fig:twr_basic} TWR timing diagram; %
    \protect\subref{fig:loc} Localization demo setup%
  }
\end{figure}

The number of users connected to a single AP is limited and, therefore, a two way ranging is used for localization. The main disadvantage of 
TWR is that it requires a large number of transmissions for localization. In dense scenarios this would lead to usage of large portion of the medium only for localization.
Anyway, the main advantage of the TWR is that it does not require synchronization between nodes, which makes it easy for implementation. 
The TWR is basically consisted of two transmissions, as shown in Figure \ref{fig:twr_basic}.

The round trip time, $t_{\text{round}}$, and the reply time, $t_{\text{reply}}$, are estimated in the nodes and the time of flight is easily calculated afterwards.

The localization is performed using trilateration. The mobile station (MS) performs TWR with multiple APs and obtains the distance. Having the distance from the MS to the four APs, the three dimensional position 
of the MS can be estimated by using trilateration. 

The overall process of localization is controlled using a localization server. The entities which want to obtain information about the location of a given node should setup the 
localization parameters using the localization server. The server issues the commands for TWR and estimates the position based on the received (estimated) distances between the 
MS and the APs.

The overall concept was tested using software defined radios (SDR). In Figure \ref{fig:loc} is shown the setup used for testing. Four SDRs are used as anchor nodes and one is used as a mobile node.
A localization presicsion and accuracy of \SI{1}{\m} is achieved using a bandwidth of \SI{25}{\MHz} in \SI{5}{\GHz} band.


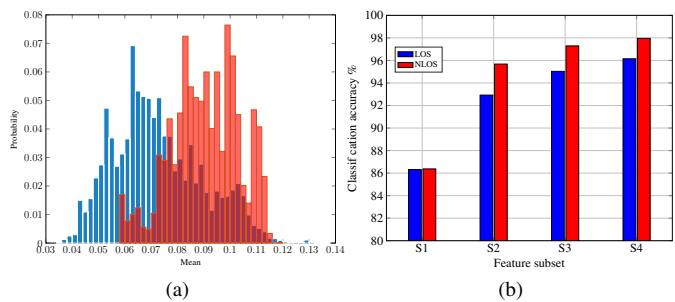
\begin{figure}[b]
  \centering
  \subfloat[]{
%
%
\definecolor{mycolor1}{rgb}{0.00000,0.44700,0.74100}%
\definecolor{mycolor2}{rgb}{0.85000,0.32500,0.09800}%
\begin{tikzpicture}
\pgfplotsset{every tick label/.append style={font=\large},scaled ticks=false, tick label style={/pgf/number format/fixed}}
\begin{axis}[%
width=4.521in,
height=3.566in,
at={(0.807in,0.648in)},
scale only axis,
xmin=0.03,
xmax=0.14,
xlabel={Mean},
ymin=0,
ymax=0.08,
ylabel={Probability},
axis background/.style={fill=white}
]
\addplot[ybar interval, fill=mycolor1, fill opacity=0.9, draw=white, area legend] table[row sep=crcr] {%
x	y\\
0.034	0.000119925646099418\\
0.036	0.00125921928404389\\
0.038	0.00251843856808779\\
0.04	0.00293817832943575\\
0.042	0.0149907057624273\\
0.044	0.0109132337950471\\
0.046	0.0155903339929244\\
0.048	0.0228458355819392\\
0.05	0.0274030101337171\\
0.052	0.0473106673862205\\
0.054	0.0368771361755711\\
0.056	0.0269233075493194\\
0.058	0.0312406308088985\\
0.06	0.0365173592372729\\
0.062	0.0691970977993644\\
0.064	0.0533069496911915\\
0.066	0.0513881393536008\\
0.068	0.050728548300054\\
0.07	0.0440127121184865\\
0.072	0.0509683995922528\\
0.074	0.0375367272291179\\
0.076	0.0374168015830185\\
0.078	0.0253043113269773\\
0.08	0.0295017089404569\\
0.082	0.0220063560592433\\
0.084	0.0344786232535828\\
0.086	0.0210469508904479\\
0.088	0.0276428614259159\\
0.09	0.0177489956227139\\
0.092	0.0115128620255442\\
0.094	0.0182286982071116\\
0.096	0.0159501109312226\\
0.098	0.0163698506925706\\
0.1	0.0185884751454098\\
0.102	0.0208670624212988\\
0.104	0.0166097019847694\\
0.106	0.00983390298015231\\
0.108	0.00623613359716976\\
0.11	0.00509683995922528\\
0.112	0.00395754632128081\\
0.114	0.00167895904539186\\
0.116	0.00155903339929244\\
0.118	0.000959405168795347\\
0.12	0.000479702584397673\\
0.122	0.000179888469149128\\
0.124	0.000299814115248546\\
0.126	0.000239851292198837\\
0.128	0.00107933081489477\\
0.13	0.000359776938298255\\
0.132	5.99628230497092e-05\\
0.134	5.99628230497092e-05\\
0.136	5.99628230497092e-05\\
0.138	5.99628230497092e-05\\
};
\addplot[ybar interval, fill=red, fill opacity=0.6, draw=mycolor2, area legend] table[row sep=crcr] {%
x	y\\
0.0580000000000001	0.0170029895366218\\
0.0600000000000001	0.00766068759342302\\
0.0620000000000001	0.0100896860986547\\
0.0640000000000001	0.0123318385650224\\
0.0660000000000001	0.00560538116591928\\
0.0680000000000001	0.0046711509715994\\
0.0700000000000001	0.0104633781763827\\
0.0720000000000001	0.0308295964125561\\
0.0740000000000001	0.0289611360239163\\
0.0760000000000001	0.0435351270553064\\
0.0780000000000001	0.0274663677130045\\
0.0800000000000001	0.0455904334828102\\
0.0820000000000001	0.0724962630792227\\
0.0840000000000001	0.054745889387145\\
0.0860000000000001	0.0510089686098655\\
0.0880000000000001	0.0497010463378176\\
0.0900000000000001	0.0599775784753363\\
0.0920000000000001	0.0401718983557549\\
0.0940000000000001	0.0599775784753363\\
0.0960000000000001	0.0321375186846039\\
0.0980000000000001	0.0764200298953662\\
0.1	0.0655829596412556\\
0.102	0.0450298953662182\\
0.104	0.0201793721973094\\
0.106	0.0140134529147982\\
0.108	0.046711509715994\\
0.11	0.0407324364723468\\
0.112	0.023355754857997\\
0.114	0.00336322869955157\\
0.116	0\\
0.118	0.000186846038863976\\
0.12	0.000186846038863976\\
};
\end{axis}
\end{tikzpicture}%
\label{fig_mean}}%
  \subfloat[]{\includegraphics[width=.49\linewidth]{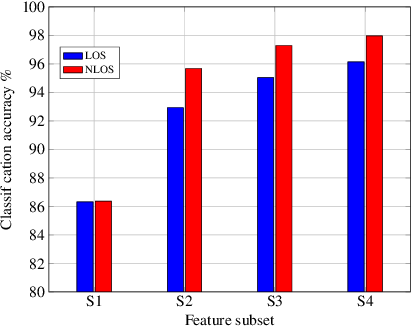}\label{perf_los_nlos}}
  \caption{
    \protect\subref{fig_mean} Mean feature. LOS in blue, NLOS in red, and the overlap in heavy red. %
        \protect\subref{perf_los_nlos} Identification performance of random forest algorithm
  }
\end{figure} 

\subsubsection*{NLOS Identification}

Non line of sight (NLOS) propagation leads to degradation in indoor wireless localization performance. When the direct path between the transmitter and the receiver is blocked, metrics such as time of arrival and angle of arrival are not correctly estimated. Therefore, approaches based on the above-mentioned metrics are expected to locate the device with an increased error margin. A reliable NLOS identification is highly recommended to achieve better indoor location and positioning accuracy.

Our proposed solution to NLOS identification problem is based on exploring features of the estimated CIR amplitude distribution from both of LOS and NLOS scenarios.

CIR estimation campaign has been carried out in an indoor office scenario. We chose features as the first to fourth central moments of channel impulse response (CIR) amplitude distribution: mean $\mu$, standard deviation  $\sigma$, skewness $s$ and kurtosis $\kappa$. Illustration of mean feature is shown in \Cref{fig_mean}.

To distinguish between LOS and NLOS scenarios, we employed the random forest (RF) machine learning algorithm \cite{Breiman2001}. The RF algorithm generates a classification model and defines some classification rules using the training dataset. These classification rules are used for prediction of new observations from the testing dataset.

We combined the best performing group of features in subsets denoted by the number of features involved, i.\,e. only one feature in subset S1 while all four features represented in subset S4 as shown in \Cref{ftsubset}. The features are extracted according to formulas in \cite{nlos_wsa}. \Cref{perf_los_nlos} depicts the identification accuracy for both LOS and NLOS for each subset of features. It is noticed that, the identification accuracy increases with the number of features used.

\begin{table}[h]
\centering
\caption{Feature Subsets}
\footnotesize
\label{ftsubset}
\begin{tabular}{|c|c|c|c|c|}
\hline
\multirow{2}{*}{Subset} & \multicolumn{4}{c|}{Feature} \\ \cline{2-5} 
                        & $\mu$ & $\sigma$ & $s$ & $\kappa$      \\ \hline
             S1           &      &  \checkmark     &    &       \\ \hline
              S2          &    &        &   \checkmark     &      \checkmark \\ \hline
			S3         &      &      \checkmark  &   \checkmark    &      \checkmark \\ \hline
               S4          &   \checkmark    &   \checkmark    & \checkmark      &  \checkmark     \\ \hline
\end{tabular}
\end{table}

\section{Conclusion}
\label{sec:conclusion}

Wireless technologies are going to significantly improve the flexibility and user experience of future-oriented industrial HMI applications such as mobile control panels or augmented reality. In this paper, we prooved, that PHY and MAC layer of WLAN system can be optimized for fitting strict industrial requirements. Furthermore, integration of localization system gains flexibility as compared to stand alone solutions.

Our future work concerns an integrated proof of concept system, which is capable of all described technologies. Furthermore, we did not yet address the aspects of mobility support. Here, multi-conne
ctivity concepts can be applied in order to provide seamless handover.

\section{Acknowledgement}

This work has been supported by the Federal Ministry of Education and Research of the Federal Republic of Germany (Foerderkennzeichen KIS3DKI018, 16KIS0245, 16KIS0249, PROWILAN). The authors alone are responsible for the content of the paper.


\bibliographystyle{IEEEtran}
\bibliography{2018-04-02_mobilkom_prowilan_radio,ref1_ihp,mobilkom_bibfile_tud}

\end{document}